\documentclass[aps,prb,twocolumn,groupedaddress]{revtex4-1}
\usepackage{amsmath,amssymb,subfigure,fancybox,txfonts,bm}
\usepackage[usenames]{color}
\usepackage[dvipdfmx]{graphicx}
\begin{document} 
\title{
Barrier effect of grain boundaries on the avalanche propagation of polycrystalline plasticity
}

\newcommand{\vc}{\mathbf}
\newcommand{\del}[2]{\frac{\partial #1}{\partial #2}}
\newcommand{\gvc}[1]{\mbox{\boldmath $#1$}}
\newcommand{\fracd}[2]{\frac{\displaystyle #1}{\displaystyle #2}}
\newcommand{\ave}[1]{\left< #1 \right>}
\newcommand{\intd}[1]{\text{d} {#1}}
\newcommand{\red}[1]{\textcolor{Red}{#1}}
\newcommand{\blue}[1]{\textcolor{Blue}{#1}}
\newcommand{\green}[1]{\textcolor[rgb]{0,0.6,0}{#1}}
\newcommand{\eng}[1]{#1}

\newcommand{\subti}[1]{}
\newcommand{\jpn}[1]{}
\newcommand{\secti}[1]{\section{#1}}
\newcommand{\niyama}[1]{\textcolor[named]{Red}{(新山) #1}}

\author{Tomoaki Niiyama}
\email{niyama@se.kanazawa-u.ac.jp}
\author{Tomotsugu Shimokawa$\dagger$}
\affiliation{
College of Science and Engineering, Kanazawa University,
Kakuma-machi, Kanazawa, Ishikawa 920-1192, Japan\\
$\dagger$Faculty of Mechanical Engineering, Kanazawa University,
Kakuma-machi, Kanazawa, Ishikawa 920-1192, Japan
}
\date{\today}
\pacs{
}

\begin{abstract}
To investigate the barrier effect of grain boundaries on
the propagation of avalanche-like plasticity at the atomic-scale,
we perform three-dimensional molecular dynamics simulations
 by using simplified polycrystal models
 including symmetric-tilt grain boundaries.
The cut-offs of stress-drop distributions following power-law distributions
decrease as the size of the crystal grains decreases.
We show that some deformation avalanches are confined by grain boundaries;
on the other hand, unignorable avalanches penetrate all the grain boundaries
included in the models.
The blocking probability that one grain boundary hinders
this system-spanning avalanche
is evaluated by using an elemental probabilistic model.
\end{abstract}

\maketitle

%

A new insight into crystalline plasticity
has been provided from non-equilibrium physics at the beginning of the century.
Discontinuous, stick-slip plastic deformation,
referred to as {\em intermittent plasticity}, 
has been revealed as an intrinsic nature of plasticity in crystalline solids
\cite{AnanthakrishnaPRE1999crossover,Miguel2001Intermittent-di,Zaiser2006IntermittentPlasticityReview,AnanthakrishnaPhysRep2007,Sethna12102007,Csikor2007DislocationAvalanche,WeissPRB2007PowerLawExpMetal,Tsekenis2013UniClassDeterm};
the probability of a deformation event
with a magnitude $s$ follows a {\em power-law distribution},
$P(s) \propto s^{-\beta}$, where $\beta$ is a constant.
This power-law distribution is a fingerprint of
the presence of non-equilibrium critical phenomena
especially self-organized criticality \cite{Bak1987SOC,SOC1998Jensen}.

A combination of acoustic emission measurements
and numerical simulations
using discrete-dislocation dynamics have revealed that
the power-law behavior of plasticity is
caused by {\em avalanches} of dislocation motions
\cite{Miguel2001Intermittent-di,Csikor2007DislocationAvalanche}.
The acoustic emission measurements in the creep of polycrystal ices
have indicated that
{\em grain boundaries (GBs)}, i.e., interfaces between crystal grains, 
can act as obstacles to the avalanches
\cite{Richeton2005Breakdown-of-av,Richeton2005DislocationAvalanchePoly}.
Louchet {\it et al.} have introduced a new concept regarding the plastic
deformation of polycrystals, which is one of the most fundamental subjects
in material science; polycrystal yielding occurs when
the avalanche transmits across GBs and percolates through the material
\cite{Louchet2006HPLaw}.
Thus, the elucidation of the interaction between GBs and
avalanches of plasticity
would advance our understanding regarding the features of
plastic deformation of polycrystals,
in particular, the grain size dependence of plastic yielding
\cite{Hall1951HPlaw,Petch1953HPlaw,Meyers2006MechProNanoCry}.

The interaction between single dislocations and GBs has been extensively
investigated \cite{Livingston1957Nvalue,Shen1988Mvalue,Shibutani2013Lvalue},
but the quantification of the interaction between avalanches of dislocations
is still quite preliminary.
The consequence of the avalanche statistics by GBs
has been discussed through discrete-dislocation dynamics
\cite{Moretti2008DDandGB}.
However, the dislocation-GB interaction 
is truly atomic-scale dynamics.
Therefore, molecular dynamics (MD) simulations for this issue
are desperately needed
to provide a correct description and quantification of the interaction.
Recently, intermittent plasticity in single crystals
has been successfully reproduced
by MD simulations \cite{moretti2011yielding,Niiyama2015ICPMD}.
However, intermittent plasticity in polycrystals remains unaddressed.

In this study, by performing MD simulations
with polycrystal models consisting of some symmetric-tilt GBs,
we prove and evaluate the role of GBs as obstacles for intermittent plasticity.
In particular, we demonstrate the statistical distribution of
the mechanical response of tensile deformation, 
its dependence on the grain size, and the atomic-scale dynamics of
avalanche motion in the polycrystal models.
Finally, we attempt to quantify
the blocking probability of avalanche propagation by GBs.

%

We performed three-dimensional MD simulations
of uniaxial tensile deformation of aluminum polycrystals 
under constant temperature and strain rate condition
with different grain sizes,
by applying the embedded atom potential presented by Mishin {\it et al.}
to the atomic interaction \cite{Mishin1999EAMAl}.
For the simulations, we employed polycrystal models
simplified as lamellar stacking structures
containing several $\ave{112}\Sigma 11$ symmetric-tilt GBs,
which align normal to the tensile direction.
Owing to this alignment,
GB sliding and grain-growth are sufficiently suppressed
in this simulation because neither process contributes
to releasing the tensile stress.
There are various types of GBs,
such as symmetric, asymmetric, and twisted \cite{Sutton1995GB};
for simplicity,
we employed symmetric-tilt GBs for initial the trials.

\begin{figure}[tbp]
 \begin{center}
  \includegraphics[width=8.5cm,bb=0 0 308 162]{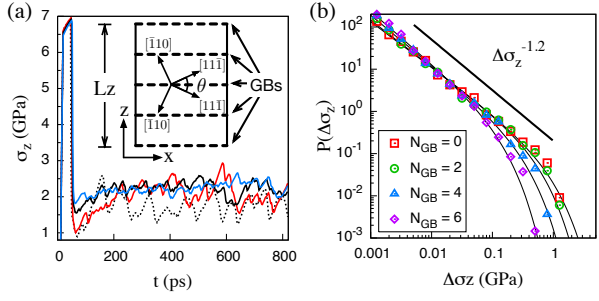}
  \caption{\footnotesize
  (a) Typical stress-time curves,
  where the tensile stress $\sigma_z(t)$ of
  $N_{GB} = 0$, $2$, $4$, and $6$ models are depicted as
  dashed black, solid red, black, and blue lines, respectively.
  The schematic figure of the polycrystal models used in this study is imposed.
  (b) The probability distributions of stress drop,
  $\Delta \sigma_z$, in each model,
  where the empirical distributions in Eq.~(\ref{eq:fit-pow-cut})
  fitted to each distribution are shown as thin black curves.
  }
  \label{fig:t-stress-dist}
 \end{center}
\end{figure}

Here we describe the procedure for preparing the polycrystal models
including $\ave{112}\Sigma 11$ symmetric-tilt GBs. 
First, let us consider two fcc lattices,
whose $[11\bar{1}]$, $[112]$, and $[1\bar{1}0]$ axes 
are along the $x$, $y$, and $z$ directions, respectively.
Next, we tilt the fcc lattices
around the $[112]$ axis by $+ \theta/2$ and $- \theta/2$, respectively.
The GB misorientation angle
$\theta/2 = \tan^{-1}\sqrt{3/8} \simeq 31.48^\circ$ is chosen.
GBs with this angle are called $\Sigma 11$ GBs
\cite{Shimokawa2010GBMD,Sutton1995GB}.
If one joins those lattices at an $xy$-plane interface,
the interface is the $\ave{112}\Sigma 11$ symmetric-tilt GB.
Both tilted lattices with the lattice constant of aluminum $a_0$
have a unit cell with minimum periodic lengths along $x$, $y$, and $z$;
$\Delta L_x = \sqrt{33/2} a_0$, $\Delta L_y = \sqrt{3/2} a_0$, and
$\Delta L_z = \sqrt{11} a_0$.
Thus, one can apply the periodic boundary conditions to the lattices
without destructing the translational symmetry, in so far as
the lengths of a simulation cell along the three directions
are equal to the integral multiple of the corresponding minimum
periodic length of the tilted lattices.
For the present simulations, we chose the dimensions of the simulation cell:
$L_x = 8 \Delta L_x \simeq 13.2$~nm, $L_y = 24 \Delta L_y \simeq 11.9$~nm,
and $L_z = 12 \Delta L_z \simeq 16.2$~nm,
where the cell is filled with $152064$ atoms.
Periodic boundary conditions were applied to the simulation cell.
In this dimension,
if the three unit cells with  $+\theta/2$ are alternately stacked with
three unit cells with $-\theta/2$ along the $z$-axis in layers,
a lamellar stacked polycrystal model is obtained with four GBs
as depicted in Fig.~\ref{fig:t-stress-dist}(a).
Following, we prepared four models with $0$, $2$, $4$, and $6$ GBs
by employing the layer thickness corresponding to the grain sizes
$12 \Delta L_z$, $6 \Delta L_z$, $3 \Delta L_z$, and $2 \Delta L_z$,
respectively.
We distinguish these models by the number of GBs, $N_{GB}$.

Using the polycrystal models as initial configurations and random
numbers as initial velocities for all atoms,
we performed MD simulations for uniaxial tensile deformation along
the $z$-axis until $50$~\% deformation is attained.
During the simulations, the pressures along the $x$- and $y$-axes
were maintained zero using the Parrinello-Rahman method without shear
\cite{Parrinello1980RahmanMethod}.
To save computing time for the simulations,
we applied a strain rate $10^{10}$~s$^{-1}$ at the early stage of
the deformation to $10$~\% when the deformation is purely elastic.
After that, the strain rate was changed to $5 \times 10^8$~s$^{-1}$.
For each model, we performed the MD simulations $20$ times
with different initial velocities for all atoms.

For the simulations, we used Langevin's dynamics represented by
the equation of atom motion \cite{tuckerman2010statistical};
\begin{equation}
 m_i \ddot{\gvc{q}}_i(t) = \gvc{F}_i - m_i \gamma \dot{\gvc{q}}_i(t) + \sqrt{2 m_i \gamma k_B T } \  \gvc{\eta}_i(t),
\end{equation}
where $m_i$ is the mass of the $i$-th atom,
$\gvc{q}_i$ and $\gvc{F}_i$ are the position and the force vector
of the $i$-th atom, respectively.
$k_B$, $T$, $\gamma$, and $\gvc{\eta}_i(t)$ are Boltzmann's constant,
temperature, the friction coefficient, 
and the random force vector as a white noise, respectively.
In the present study, we chose $\gamma = 1$~ps$^{-1}$.
The selected temperature was $10$~K,
to reduce thermal fluctuations \cite{Niiyama2015ICPMD}.
This equation of motion is also employed to the dynamics
of the simulation cell.
The second-order accurate algorithm developed
by Vanden-Eijinden and Ciccotti was used
for the numerical integration of the dynamics
\cite{VandenEijnden2006310,tuckerman2010statistical}.
To generate the random forces, Mersenne Twister generator was employed
\cite{Matsumoto1998MT}.

The noise-induced dynamics not only keeps the temperature constant
but also depresses stress fluctuations irrelevant to plastic deformation;
the fluctuations are caused by phonon propagation
or periodic dislocation motion such as string vibration
after large-scale plastic deformation events.


The tensile stress of all the models, $\sigma_z(t)$,
obtained from the simulation results
shows intermittent plastic manners
as depicted in Fig.~\ref{fig:t-stress-dist}(a);
$\sigma_z(t)$ gradually increases by elastic deformation and
abruptly drops by repeated plastic deformation,
where the time-series are averaged over a $0.2$~ps interval
to remove thermal fluctuations.
This serrated behavior is the same as the intermittent plasticity
observed in previous numerical studies
\cite{moretti2011yielding,Niiyama2015ICPMD}, 
but the amplitude of the stress fluctuation decreases as 
$N_{GB}$ increases.


The stress-drop distribution extracted from the average time-series
are shown in Fig.~\ref{fig:t-stress-dist}(b).
The value of stress drop, $\Delta \sigma_z$, is defined
as the reduction amount of
$\sigma_z$ during {\em a plastic deformation event}
from $t_{start}$ to $t_{end}$; 
$\Delta \sigma_z = \sigma_z(t_{start}) - \sigma_z(t_{end})$.
Plastic deformation events can be determined as periods
during which $\sigma_z(t)$ monotonically decreases
as explained in the previous study \cite{Niiyama2015ICPMD}.
In the figure,
all the distributions indicate algebraic decay in the range of
$\Delta \sigma_z < 0.1$~GPa.
This feature corresponds to a power-law behavior, which is consistent with
the results obtained in previous numerical studies
employing MD \cite{moretti2011yielding,Niiyama2015ICPMD}
and discrete-dislocation dynamics simulations
\cite{Csikor2007DislocationAvalanche,Miguel2001Intermittent-di,Tsekenis2013UniClassDeterm},
but there is a notable difference.
That is, the distributions have different {\em cut-offs}, which are
rapid decreases of probability at a large-scale regime.
We evaluated the cut-offs of the distributions
by fitting the following empirical relationship
to each distribution:
\begin{equation}
 P(\Delta \sigma_z) \approx
  {\Delta \sigma_z}^{-\beta} \exp(-\Delta \sigma_z/\Delta \sigma_c),
\label{eq:fit-pow-cut}
\end{equation}
where $\beta$ and $\Delta \sigma_c$ are the power-law exponent
and the characteristic scale of stress drops, respectively.
The values of $\beta$ and $\Delta \sigma_c$ obtained by the fitting
are listed in Table~\ref{table}.
As can be seen from the table and Fig.~\ref{fig:t-stress-dist}(b),
$\Delta \sigma_c$ decreases as the number of GBs in each model increases,
whereas the exponents of the models are almost the same ($\beta \simeq 1.2$).
This trend of the cut-offs supports that our MD simulations reproduce
the role of GBs as obstacles to the avalanches consistent with
the previous experimental study
\cite{Richeton2005Breakdown-of-av}.

 \begin{table}[tbp]
  \caption{\footnotesize
  The values of the power-law exponent $\beta$,
  the cut-offs $\Delta \sigma_{c}$, and the frequency of occurrence
  of system-spanning deformation events $n_{SS}$ and
  all deformation events $n_{def}$.
  }
  \label{table}
 \begin{tabular}{p{1 cm}  p{1.5 cm} p{1.5 cm}  p{1.5 cm} p{1.5 cm}}
  \hline
  \hline
  $N_{GB}$  & $\beta$ & $\Delta \sigma_{c}$ & $n_{SS}$ & $n_{def}$\\
            &         &       (GPa)       &   \\
  \hline
   0        & 1.11    &  0.788  & 142 & 644\\
   2        & 1.08    &  0.480  & 123 & 694\\
   4        & 1.19    &  0.288  & 126 & 870\\
   6        & 1.24    &  0.121  & 126 & 1080\\
  \hline
  \hline
 \end{tabular}
 \end{table}


In addition to the role of GBs
revealed from the macroscopic responses,
MD simulations can allow us to confirm this behavior
directly by observing microscopic-scale dynamics
during the deformation process.
Here we analyze statistically the linear size
of a slip area, during a plastic deformation event,
by identifying slip areas in the following way.

Let us consider the set of $12$ nearest neighbor atoms around the $i$-th atom
at time $t$.
We denote these atoms by
$\gvc{n}^{(i)}(t) = \left\{ n_1, n_2, ..., n_{12} \right\}$,
where $n^{(i)}_1, n^{(i)}_2, ..., n^{(i)}_{12}$
are integers which identify the neighbor atoms around the $i$-th atom.
If the set of the atoms at $t_{start}$ is different from that at $t_{end}$,
i.e., $\gvc{n}^{(i)}(t_{start}) \ne \gvc{n}^{(i)}(t_{end})$,
the $i$-th atom is regarded as a participant atom to the deformation event
starting at $t_{start}$.
Applying this procedure to all the atoms at a deformation event
one can identify the participant atoms that represent the slip area
caused by the deformation event.
By following this procedure,
we extract all the participant atoms from the numerical results
that are averaged with $1$~ps intervals.

\begin{figure}[tbp]
 \begin{center}
  \includegraphics[width=8cm]{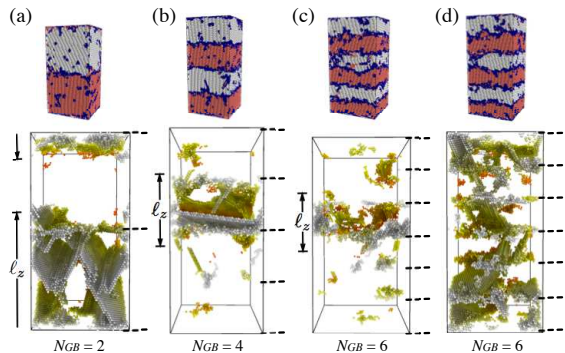}
  \caption{\footnotesize
  Snapshots of all atoms (top) and
  participant atoms representing slip areas during a plastic
  deformation event (bottom)
  in (a) the model with $N_{GB} = 2$, (b) $N_{GB} = 4$,
  (c, d) $N_{GB} = 6$.
  The approximate positions of GBs are indicated by horizontal dashed lines.
  (d) The slip area percolates all the GBs in the model.
    }
  \label{fig:snapshots}
 \end{center}
\end{figure}

Typical configurations of participant atoms identified by the above procedure
are depicted in the bottom panels of Fig.~\ref{fig:snapshots},
where smaller clusters consisting of less than $12$ participant atoms
are removed for the visibility.
The participant atoms formed into sheets as shown in the panels
follow the trail of dislocations.
From the bottom panels of Figs.~\ref{fig:snapshots}(a), (b), and (c),
it can be directly confirmed that GBs confine the propagation of an avalanche.
However, not all avalanches are blocked by GBs.
For instance, Fig.\ref{fig:snapshots}(d) shows
{\em a system-spanning deformation event}
in which a slip area penetrates all the GBs in the system,
even the system contains six GBs.
Thus, the statistics of the spatial extension of the avalanches
should be investigated quantitatively.

Note that we confirm neither fracture nor crack nucleation
in the simulations owing to the high ductility of aluminum.
For instance, the upper panels of Fig.~\ref{fig:snapshots} show
snapshots of all the atoms at the same moments as in the bottom panels,
where atoms with a defect-type lattice structure are colored blue,
and fcc structure atoms are colored red and gray
in accordance with their orientation angles.
The snapshots clearly indicate neither cracks nor fracture.


Here we evaluate the linear size of one avalanche along the $z$-axis
by the following.
The cluster analysis is performed to participant atoms
during a deformation event; 
participant atoms whose distance between each other is
$< 1.2 a_0/\sqrt{2}$ are regarded
as a member of one common cluster consisting of an isolated slip area.
The length along the $z$-axis of the largest one of all
the clusters during the event is denoted by $\ell_z$.
We represent the relative propagation length of an avalanche
as $S_z \equiv \ell_z/L_z$.
For instance, $S_z$ will be $1/N_{GB}$ in the model containing $N_{GB}$ GBs
when an avalanche emerging from a crystal grain 
is completely blocked by the neighboring GBs.
On the other hand, if an avalanche percolates through the system,
one can obtain $S_z = 1$.

\begin{figure}[tbp]
  \centering
  \includegraphics[width=8cm,bb=0 0 189 133]{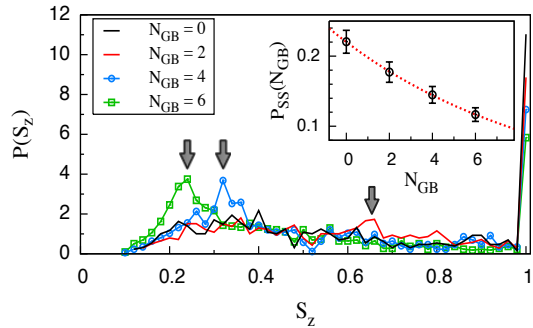}
  \caption{\footnotesize
    The distributions of the relative linear size of an avalanche $S_z$.
    The inset shows the probability of system-spanning events plotted
    against the number of GBs, $N_{GB}$.
    Open circles and the red dotted line indicate
    the probability estimated from the simulation results in Table \ref{table}
    and the theoretical relationship depicted
    in Eq.(\ref{eq:P_SS}), respectively.
 }
  \label{fig:dist-Sz}
\end{figure}

The statistical distributions $P(S_z)$ calculated from all the configurations
of participant atoms are shown in Fig.~\ref{fig:dist-Sz}.
One can find typical peaks
(indicated by arrows) at
$S_z = 0.67$, $0.33$, and $0.25$ for $N_{GB} = 2$, $4$, and $6$, respectively,
even though the peak at $S_z = 0.67$ is vague.
These are only slightly larger than the
corresponding characteristic lengths of a crystal grain in each model,
$1/N_{GB}$.
This agreement clearly indicates an effect of GBs to block the propagation
of avalanches of plasticity.
The fact that there are no typical peaks in the model with $N_{GB} = 0$
also supports the presence of the effect indirectly.
In contrast to the typical peaks providing evidence of the barrier effect,
all the distributions have significant
peaks at $S_z = 1$, which corresponds to system-spanning deformation events
as depicted in Fig.~\ref{fig:snapshots}(d), for instance.
These significant peaks mean that not a few avalanches pass through all GBs.
Because the linear size of such a system-spanning avalanche event
cannot be defined,
evaluating the mean free-path of the avalanches is impossible.
Thus, it is also impossible to quantify the barrier effect of GBs
by assessing the mean free-path.


To overcome this difficulty, we focus on the system-spanning events
rather than on all the deformation events.
We now evaluate {\em the blocking probability};
the probability that one GB hinders a system-spanning avalanche.
In other words, the probability that a deformation event evolves to
a system-spanning event is considered.
This probability can be estimated by the relative frequency of
the system-spanning events;
$P_{SS}(N_{GB}) = n_{SS} / n_{def}$, where
$n_{SS}$ and $n_{def}$ are the frequency of the system-spanning avalanche events
and that of all deformation events, respectively.
In Table~\ref{table}, we enumerate $n_{SS}$ and $n_{def}$ obtained
from the present simulations.
The inset in Fig.~\ref{fig:dist-Sz} shows
the system-spanning probability estimated from the data
plotted by open circles as a function of $N_{GB}$,
where the error bars are approximately calculated from a standard deviation
of the binomial distribution with the number of trials $n_{def}$
and the probability in each trial $n_{SS}/n_{def}$.

The estimated probability in Fig.~\ref{fig:dist-Sz}
monotonically decreases with the increase of $N_{GB}$.
In other words, the propagation of an avalanche of plasticity
is quickly damped as the grain size (grain thickness) is small.
Thus, the present result is direct evidence of the barrier effect of GBs
upon the avalanche propagation.


If one GB decreases $P_{SS}$ independently by the factor $\alpha\ (\le 1)$,
the probability in the model including $N_{GB}$ GBs
can be described by the simple relationship:
\begin{align}
 P_{SS}(N_{GB}) = P_{SS}(0) \ \alpha^{N_{GB}},
 \label{eq:P_SS}
\end{align}
where $1 - \alpha$ corresponds to
the blocking probability of system-spanning avalanches by one GB.
This theoretical relationship can fit well with
the estimated probability by the least squares method with
the parameters $P_{SS}(0) = 0.22$ and $\alpha = 0.90$
as shown in Fig.~\ref{fig:dist-Sz}.


The theoretical relationship of Eq.~(\ref{eq:P_SS}) is
made on the basis of two assumptions;
(i) the emergence of the avalanches is independent of the presence of GBs,
and (ii) a GB blocks a system-spanning avalanche independently
with the blocking probability $1-\alpha$.
While these assumptions might be disputable,
there is collateral evidence of the former assumption.
That is,
the power-law exponent of the stress-drop distribution $\beta$
does not depend on $N_{GB}$ 
as shown in Fig.~\ref{fig:t-stress-dist}(b) and Table~\ref{table}.
This trend implies that avalanches occur in the same fashion
inside grains.
The excellent agreement between the theoretical curve and the estimated
probability (the inset in Fig.~\ref{fig:dist-Sz}) supports the assumption (ii).
Hence, we can conclude that this fitting can provide us
with an estimated value of the barrier capability of the GB against
the avalanche of crystalline plasticity.


An unchanged of $\beta$ is
inconsistent with the previous experimental observations
\cite{Richeton2005Breakdown-of-av},
whereas the change of $\Delta \sigma_c$ is consistent with them.
This inconsistency might be because of the simplification of
the present polycrystal models.
That is, the present result intimates that modulation of $\beta$
observed in the previous experiment \cite{Richeton2005Breakdown-of-av}
results from complicated domain structures,
variation in the size of grains,
or both as is common in real polycrystals.
This complexity might produce some slow relaxation processes such as
GB sliding or grain-growth, which
can affect the avalanche properties
\cite{Papanikolaou2012Quasi-periodic,Cahn20064953}.
The presence of surface might also be important for the avalanche
dynamics\cite{Po2014DDDprogress,Papanikolaou2015DDDpillar}.
MD simulations employing such realistic polycrystal models
are required to prove the cause of the modulation.

To evaluate the barrier effect of a particular kind of GBs,
simplified polycrystal models similar to those used in this study
is appropriate rather than realistic polycrystal models.
It is expected that the factor $\alpha$ depends on
the misorientation angle of the tilt-and-twist GBs.
Investigation of the barrier effect of these GBs
is an issue for future studies.

The quantification of the barrier effect in this study
will contribute to the construction of theoretical models or
semi-macroscopic numerical models,
such as models for the discrete-dislocation dynamics 
or the phase field method for polycrystalline solids, which worked well
for the investigation of the avalanche behaviors in single crystals
\cite{Miguel2001Intermittent-di,Koslowski2004SOCphasefield,Csikor2007DislocationAvalanche}.
These models will shed light on the mechanical properties
of polycrystalline materials from the viewpoint of intermittent plasticity.


In this study, to investigate the interaction between
the avalanche-like propagation of plasticity and the grain boundaries (GBs)
in polycrystalline solids at the atomic-scale,
we performed molecular dynamics simulations for uniaxial tensile deformation
in aluminum polycrystal models simplified as lamellar stacking structure
including symmetric-tilt GBs.
The results show that the stress drops caused
by an avalanche of plastic deformation
follow power-law distributions even in polycrystals,
but the cut-off of the distribution decreases with decreasing grain size.
By observing atomic-scale dynamics of the simulation results, 
we noted that some avalanches are confined by GBs,
but others transmit across the GBs
or penetrate through the system entirely (system-spanning avalanches).
We propose the theoretical description of the system-spanning probability
in the models.
The excellent agreement between the theoretical probability
and the estimated probability from the simulations
provides us with the blocking probability of the avalanches
by single-grain boundary.

\begin{acknowledgments}
This research was supported by the Ministry of Education, 
Culture, Sports, Science and Technology (MEXT) 
KAKENHI Grant Number 22102007 and 16K17764, and the Japan Science and 
Technology Agency (JST) under Collaborative Research Based on 
Industrial Demand ``Heterogeneous Structure Control: 
Towards Innovative Development of Metallic Structural Materials''.
\end{acknowledgments}

\bibliographystyle{apsrev}

\end{document}